\documentclass[conference]{IEEEtran}
\usepackage{amsmath,amsfonts}
\usepackage{graphicx,amssymb}
\usepackage{algorithmic,algorithm}
\usepackage{amsthm,dsfont}
\usepackage{subfigure,url}
\usepackage[noadjust]{cite}
\usepackage{xcolor}
\usepackage{colortbl,booktabs}

\theoremstyle{plain}

\theoremstyle{remark}

\newcounter{longequ}[longequ]
\addtocounter{longequ}{11}

\allowdisplaybreaks

\IEEEoverridecommandlockouts\IEEEpubid{\makebox[\columnwidth]{ 978-1-6654-3540-6/22~\copyright~2022 IEEE \hfill} \hspace{\columnsep}\makebox[\columnwidth]{ }}
\begin{document}
%
\title{Augmented Deep Unfolding for Downlink Beamforming in Multi-cell Massive MIMO With Limited Feedback
}
\author{\IEEEauthorblockN{Yifan Ma\IEEEauthorrefmark{1},
		Xianghao Yu\IEEEauthorrefmark{1}, Jun Zhang\IEEEauthorrefmark{1}, S.H. Song\IEEEauthorrefmark{1}\IEEEauthorrefmark{2}, and Khaled B. Letaief\IEEEauthorrefmark{1}}\\
	\IEEEauthorblockA{\IEEEauthorrefmark{1}Dept. of ECE, The Hong Kong University of Science and Technology, Hong Kong\\
	\IEEEauthorrefmark{2}Division of ISD, The Hong Kong University of Science and Technology, Hong Kong\\}
	Email: \IEEEauthorrefmark{1}\{ymabj, eexyu, eejzhang, eeshsong, eekhaled\}@ust.hk
\thanks{This work was supported by the Hong Kong Research Grants Council under Grant No. 16212120 and 15207220.}}


%


\maketitle

\begin{abstract}
In limited feedback multi-user multiple-input multiple-output (MU-MIMO) cellular networks, users send quantized information about the channel conditions to the associated base station (BS) for downlink beamforming.
However, channel quantization and beamforming have been treated as two separate tasks conventionally, which makes it difficult to achieve global system optimality. In this paper, we propose an augmented deep unfolding (ADU) approach that jointly optimizes the beamforming scheme at the BSs and the channel quantization scheme at the users. 
In particular, the classic WMMSE beamformer is unrolled and a deep neural network (DNN) is leveraged to pre-process its input to enhance the performance.
The variational information bottleneck technique is adopted to further improve the performance when the feedback capacity is strictly restricted.
Simulation results demonstrate that the proposed ADU method outperforms all the benchmark schemes in terms of the system average rate.
\end{abstract}

\IEEEpeerreviewmaketitle
\section{Introduction}
Among various enabling physical layer technologies for 5G and beyond networks, network densification and massive multiple-input multiple-output (MIMO) play critical roles in achieving ultra-high spectral efficiency \cite{boccardi2014five, KBL226G}. 
In the hotspot of 5G cellular coverage scenarios such as transport hubs, stadiums, and shopping malls, multi-cell multi-user MIMO (MU-MIMO) systems are deployed to support ubiquitous high-speed data transmission.
In these systems, users may suffer from both intra- and inter-cell interference and it is essential to identify an optimal beamformer to mitigate the co-channel interference \cite{Jun09TWC}.
However, conventional beamforming schemes usually assume perfect channel state information (CSI) available at the base station (BS), which is unrealistic due to exceedingly high feedback overhead. 

There have been many works that investigated beamforming design for MIMO systems with limited feedback \cite{Vincent14CS,sohrabi21deep, Guo21Cellular}. Compressive sensing (CS) \cite{HDCSI} has been widely adopted to recover the sparse channel parameters and subsequently feed back the quantized version of these parameters to the BS \cite{Vincent14CS}. To design the beamforming matrix, the BS first reconstructs the channels and then employs a classic beamforming scheme, e.g., the zero forcing (ZF) or weighted minimum mean square error (WMMSE) algorithm \cite{Shi11WMMSE}. However, separately considering CSI quantization and beamforming makes it hard to achieve global optimality. 

With recent successes of deep learning, there has been a growing interest in developing data-driven, and in particular deep neural network (DNN)-based methods for \emph{end-to-end design}, where distributed quantization, feedback, and multi-user beamforming are jointly considered by training a DNN at each user and a DNN at the BS, respectively \cite{sohrabi21deep, Guo21Cellular}. 
However, in these methods, conventional signal processing modules are treated as a black-box and replaced by standard neural network architectures \cite{Yifan21NC}. 
In this way, the DNN-based end-to-end design does not take into account the unique characteristics of specific wireless systems, making it difficult to extract the underlying high-dimensional mapping, especially when the network size increases. 
To overcome such drawbacks, deep unfolding was recently proposed, which incorporates the domain knowledge derived from iterative optimizaton algorithms \cite{He20Model}. It unrolls the iterative algorithm, regards each iteration as one layer of the neural network, and introduces a number of trainable parameters in each layer to enhance system performance. 
For example, in \cite{Hu21Unfold}, a deep unfolding neural network based on the structure of the WMMSE algorithm was developed for beamforming in massive MIMO systems. Nevertheless, it is challenging to implement deep unfolding methods in end-to-end design. In particular, deep unfolding is typically developed based on classic optimization algorithms, which are hardly available for complicated end-to-end design due to the lack of carefully-designed models and rigorous mathematical derivation. Furthermore, while one can simply cascade several deep unfolding approaches, it will suffer from high difficulty in training such a prohibitively large number of trainable parameters.

To overcome these drawbacks, in this paper, we propose an augmented deep unfolding (ADU) method to enable the end-to-end design of the channel quantization and downlink beamforming in limited feedback MU-MIMO cellular systems. Distributed DNNs are deployed at the user side for quantization while a DNN at the BS is designed to augment the deep unfolded beamforming algorithm. Specifically, given the perfect downlink CSI at the user side, one encoding DNN is reused by different users to perform CSI quantization. At the BS side, instead of unfolding the iterative method and introducing a large number of trainable parameters at each layer, the function derived from an existing finite-iteration method, i.e., the WMMSE beamformer, is kept intact and a DNN is applied to pre-process the input of this function. Furthermore, to improve the performance when the feedback bits are limited, a better trade-off between the ultimate system performance and the communication overhead has to be identified, for which the information bottleneck (IB) framework is adopted. 
In particular, the mutual information between the feedback bits and the input channels is minimized during training and the variational approximation, i.e., variational information bottleneck (VIB), is adopted to derive a tractable upper bound for the calculation of the mutual information. It is shown that by bypassing the explicit channel reconstruction stage, exploiting the domain knowledge inherent in the iterative algorithm, and leveraging the VIB technique, the resulting ADU design provides a better performance compared to both the conventional block-by-block solutions and the state-of-the-art fully data-driven design, especially when the number of feedback bits is limited and users are densely distributed. 

\section{System Model and Problem Formulation}\label{sec:sys}

\begin{figure}[t] 
\centering
\includegraphics[width=0.45\textwidth]{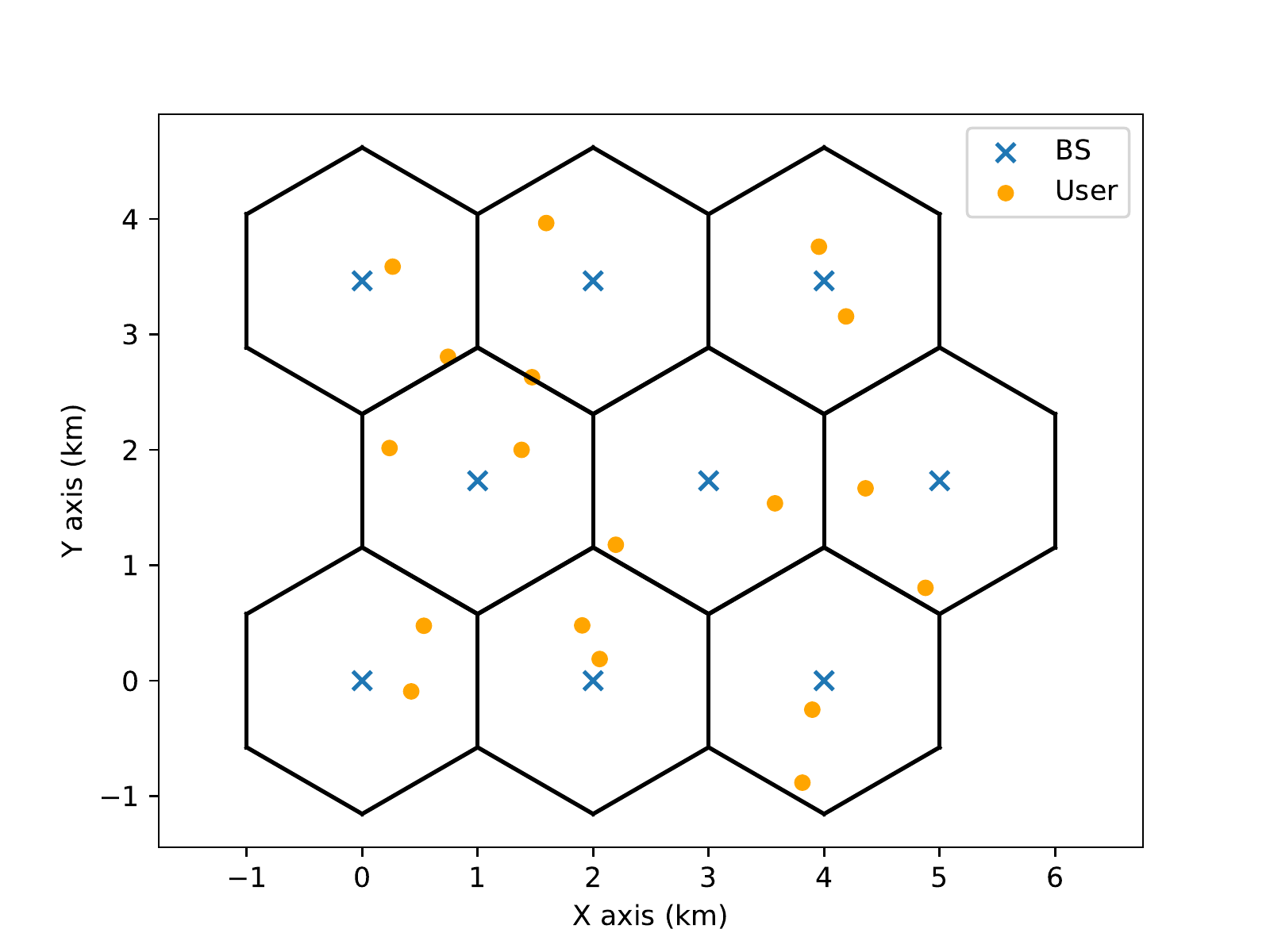} 
\caption{An illustrative example of a multi-user cellular network with $M = 9$ and $N = 18$.} 
\label{Cellular} 
\end{figure}

As illustrated in Fig. \ref{Cellular}, an $M$-cell MU-MIMO system is considered, where the $i$-th BS is equipped with $N_\mathrm{t}$ transmit antennas and serves $I_i$ users in cell $i$. Let $i_k$ be the $k$-th user in the $i$-th cell and each user has $N_\mathrm{r}$ receive antennas. Set $\mathcal{I}$ denotes the set of all receivers, i.e., 
\begin{equation}
\mathcal{I}=\left\{i_{k} \mid i \in \mathcal{M} \triangleq\{1, \ldots, M\}, k \in\left\{1,2, \ldots, I_{i}\right\}\right\}.
\end{equation}
There are in total $|\mathcal{I}| \triangleq N$ users in this multi-cell MU-MIMO systems. The channel matrix from the $j$-th BS to the $k$-th user in cell $i$ is denoted by $\mathbf{H}_{i_k, j} \in \mathbb{C}^{N_\mathrm{r} \times N_\mathrm{t}}$, where $j \in \mathcal{M}$ and $i_k \in \mathcal{I}$.

The frequency division duplexing (FDD) is assumed in this system. 
Assume that perfect downlink CSI is known at the user side. Each user transmits $B$ bits of information to its associated BS for multi-user downlink beamforming. The feedback bits for the $k$-th user in the $i$-th cell are denoted by ${\mathbf{q}}_{i_{k}} \in \mathbb{C}^{B \times 1}$.
At the BS side, each BS $i \in \mathcal{M}$ collects the feedback bits from all $I_i$ users ${\mathbf{q}}_{i} = [{\mathbf{q}}_{i_1}^T, \cdots, {\mathbf{q}}_{i_{I_i}}^T]^T \in \{\pm1\}^{I_i B}$ and design the beamforming matrix given those feedback bits. 
Let ${\mathbf{s}}_{i_k}$ denote the transmit signal vector from the $i$-th BS to the $i_k$-th user and assume that ${\mathbf{s}}_{i_k}$ is with zero mean and $\mathbb{E}{[{\mathbf{s}}_{i_k} {\mathbf{s}}_{i_k}^H]} = {\mathbf{I}}$.
Prior to transmission, the $i$-th BS linearly precodes its signal vector as $\mathbf{x}_i = \sum_{k=1}^{I_i} \mathbf{V}_{i_k} {\mathbf{s}}_{i_k}$, where $\mathbf{V}_{i_k} \in {\mathbb{C}}^{N_t \times N_r}$ denotes the beamforming matrix at the $i$-th BS to transmit the signal ${\mathbf{s}}_{i_k}$ to receiver $i_k$. 
The transmit power constraint is given by $\sum_{k=1}^{I_i} \operatorname{Tr}(\mathbf{V}_{i_k} {\mathbf{V}}_{i_k}^H) \leq P_\mathrm{T}$, where $ P_\mathrm{T}$ is the maximum downlink transmit power for the BS. Accordingly, the received signal vector $y_{i_k} \in {\mathbb{C}}^{N_r \times 1}$ at user $i_k \in \mathcal{I}$ can be written as
\begin{equation}
\begin{aligned}
\mathbf{y}_{i_{k}}=& \underbrace{\mathbf{H}_{i_{k}, i} \mathbf{V}_{i_{k}} {\mathbf{s}}_{i_{k}}}_{\text {desired signal }}+\underbrace{\sum_{m=1, m \neq k}^{I_{i}} \mathbf{H}_{i_{k}, i} \mathbf{V}_{i_{m}} {\mathbf{s}}_{i_{m}}}_{\text {intra-cell interference }} \\
&+\underbrace{\sum_{j=1, j \neq i}^{M} \sum_{\ell=1}^{I_{j}} \mathbf{H}_{i_{k}, j} \mathbf{V}_{j_{\ell}} {\mathbf{s}}_{j_\ell}}_{\text {inter-cell interference }}+\mathbf{n}_{i_{k}},
\end{aligned}
\end{equation}
where $\mathbf{n}_{i_k} \in \mathbb{C}^{N_r\times1}$ represents the additive white Gaussian noise (AWGN) at the $i_k$-th user with $\mathbb{E}{[\mathbf{n}_{i_k} \mathbf{n}^H_{i_k}]} = \sigma^2_{i_k} \mathbf{I}$.

This paper aims to jointly design the channel quantization scheme at the users and beamforming scheme at the BSs to maximize the system sum-rate. The problem formulation is given by
\begin{equation}
\begin{aligned}
\label{main_problem}
\max_{\mathcal{F}(\cdot), \mathcal{G}(\cdot)} ~ & \sum_{i=1}^{M}\sum_{k=1}^{I_i} R_{i_k} = \sum_{i=1}^{M}\sum_{k=1}^{I_i} \log \operatorname{det}(\mathbf{I}+\mathbf{H}_{i_{k}, i} \mathbf{V}_{i_{k}} \mathbf{V}_{i_{k}}^{H} \mathbf{H}_{i_{k}, i}^{H} \\
&(\sum_{m=1, m \neq k}^{I_{i}} \mathbf{H}_{i_{k}, i} \mathbf{V}_{i_{m}} \mathbf{V}_{i_{m}}^{H} \mathbf{H}_{i_{k}, i}^{H}+ \\
&\sum_{j=1, j \neq i}^{M} \sum_{\ell = 1}^{I_j} \mathbf{H}_{i_{k}, j} \mathbf{V}_{j_{\ell}} \mathbf{V}_{j_{\ell}}^{H} \mathbf{H}_{i_{k}, j}^{H}+\sigma_{k}^{2} \mathbf{I})^{-1})\\ 
\text{s.t.} ~ & {\mathbf{q}}_{i_{k}}=\mathcal{F}\left(\mathbf{H}_{i_{k}, i}\right), ~  \forall i_k \in \mathcal{I},\forall i \in \mathcal{M}, \\
& \mathbf{V}_{i_{k}}=\mathcal{G}\left({\mathbf{q}}_{i}\right), ~  \forall i_k \in \mathcal{I}, \forall i \in \mathcal{M},
\end{aligned}
\end{equation}
where the function $\mathcal{F}(\cdot): \mathbb{C}^{N_r \times N_t} \rightarrow \{\pm 1\}^B$ represents the feedback scheme adopted at the $i_k$-th user and the function $\mathcal{G}(\cdot) : \{\pm 1\} ^{I_i B} \rightarrow \mathbb{C}^{N_t \times N_r}$ represents the downlink beamforming scheme adopted at the $i$-th BS. Note that for the considered end-to-end design, the channel quantization scheme $\mathcal{F}(\cdot)$ and the downlink beamforming scheme $\mathcal{G}(\cdot)$ are jointly optimized and the explicit downlink channel reconstruction phase at the BSs is bypassed. In the fully data-driven end-to-end method \cite{sohrabi21deep, Guo21Cellular}, these two mappings can be approximated by DNNs trained with a large amount of data.
However, since fully data-driven methods treat conventional signal processing modules as a black-box without incorporating any domain knowledge, the underlying high-dimensional mapping is difficult to learn when users are densely distributed \cite{Yifan21NC}. Therefore, it is difficult for these methods to achieve satisfactory performance in multi-cell MU-MIMO networks due to severe interferences.

\section{Proposed Augmented Deep Unfolding}
In this section, we introduce the ADU-based method for solving Problem \eqref{main_problem}, which combines advantages of both the classic iterative optimization method and data-driven method. 
\subsection{Architecture of the Proposed Augmented Deep Unfolding}
Note that the downlink beamforming for multi-cell MU-MIMO systems is NP-hard even if perfect downlink CSI at the BS (CSIT) is available. 
One of the well-acknowledged algorithms in literature is WMMSE \cite{Shi11WMMSE}, whose iterative updating rule is given by 
\begin{equation}\label{WMMSE}
\begin{aligned}
&\mathbf{U}_{i_k}=\mathbf{A}_{i_k}^{-1} \mathbf{H}_{i_k, i} \mathbf{V}_{i_k},\\
&\mathbf{W}_{i_k}=\mathbf{E}_{i_k}^{-1},\\
&\mathbf{V}_{i_k}= \mathbf{B}^{-1} \mathbf{H}_{i_k, i}^{H} \mathbf{U}_{i_k} \mathbf{W}_{i_k},
\end{aligned}
\end{equation}
where $\mathbf{A}_{i_k}= \frac{\sigma_{i_k}^{2}}{P_\mathrm{T}} \sum_{i_k=1}^{N} \operatorname{Tr}\left(\mathbf{V}_{i_k} \mathbf{V}_{i_k}^{H}\right) \mathbf{I}+\sum_{j=1, j \neq i}^{M} \sum_{\ell=1}^{I_{j}} \mathbf{H}_{i_k, j} \mathbf{V}_{j_\ell} \mathbf{V}_{j_\ell}^{H} \mathbf{H}_{i_k, j}^{H}$, $\mathbf{E}_{i_k}=\mathbf{I}-\mathbf{U}_{i_k}^{H} \mathbf{H}_{i_k, i} \mathbf{V}_{i_k}$, and $\mathbf{B}=\sum_{k=I_1}^{I_i} \frac{\sigma_{i_k}^{2}}{P_\mathrm{T}} \operatorname{Tr}\left( \mathbf{U}_{i_k} \mathbf{W}_{i_k} \mathbf{U}_{i_k}^{H}\right) \mathbf{I}+\sum_{m=1}^{I_i} \mathbf{H}_{i_m}^{H} \mathbf{U}_{i_m} \mathbf{W}_{i_m} \mathbf{U}_{i_m}^{H} \mathbf{H}_{i_m}$. For a fixed iteration WMMSE method, let $f_{\mathrm{WMMSE}}(\cdot)$ denote the mapping from the downlink channel $\mathbf{H}_{i_k, j}$ to the optimized beamforming solution $\mathbf{V}_{i_k}$, where $j \in \mathcal{M}$ and $i_k \in \mathcal{I}$. Although excellent performance of the WMMSE algorithm has been observed experimentally and theoretically, implementing it in real systems still faces many serious obstacles. On the one hand, the high computational cost incurred by WMMSE, e.g., matrix inverse operation and large number of iterations, defers the real-time implementation in practical systems. On the other hand, numerical optimization algorithms do not support end-to-end design and may suffer from performance loss because of the block-by-block structure. Since iterative optimization algorithms are typically based on carefully-designed model and rigorous mathematical derivation, for the complicated joint design Problem \eqref{main_problem}, it is challenging to directly identify an effective numerical optimization algorithm. 

\begin{figure}[t] 
\centering
\includegraphics[width=0.5\textwidth]{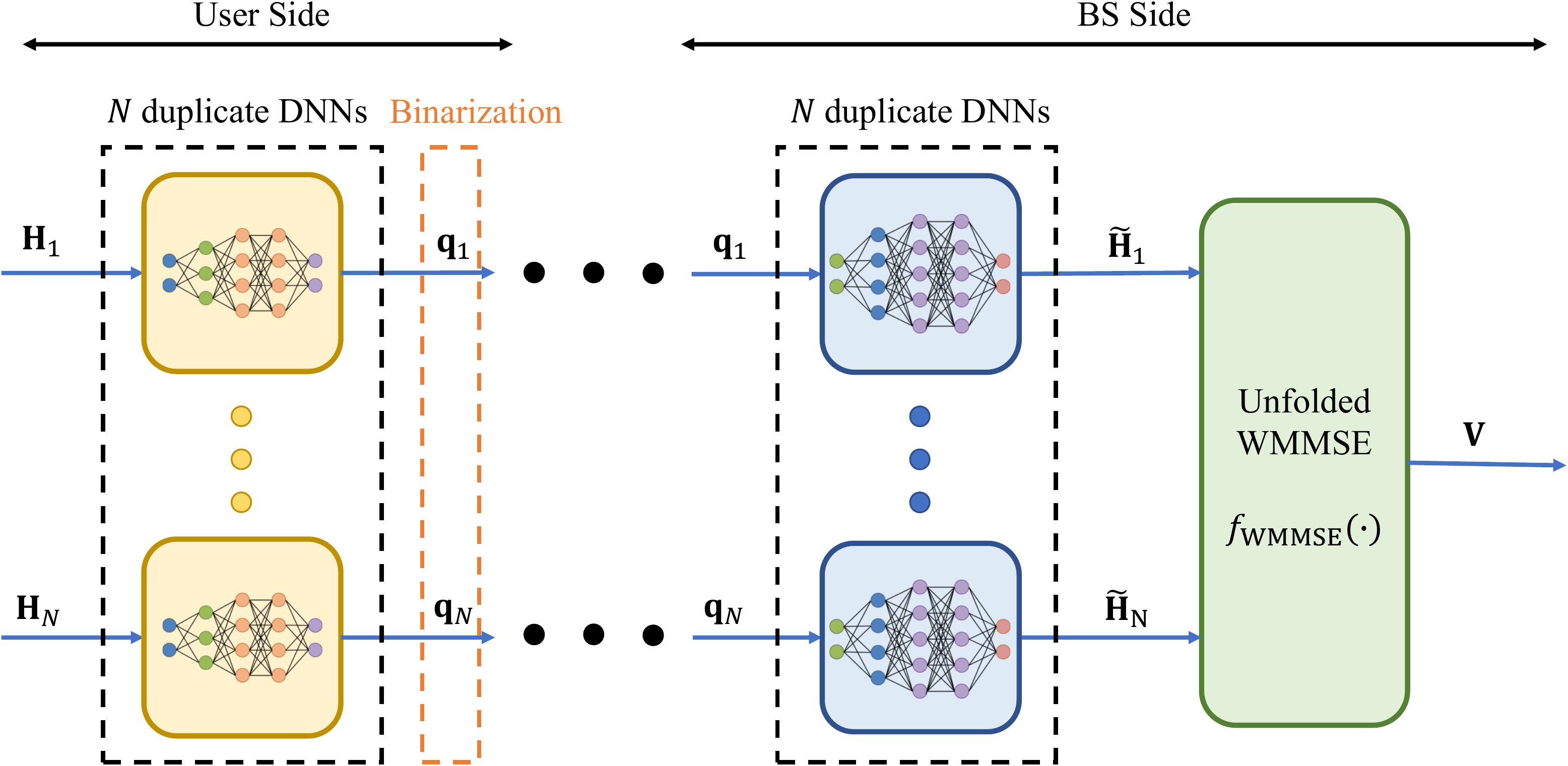} 
\caption{Proposed ADU-based end-to-end design architecture for FDD MU-MIMO cellular networks.} 
\label{architecture} 
\end{figure}

To address these problems, we propose an ADU method that integrates iterative algorithms with deep learning. Different from existing deep unfolding methods that unroll the iterative optimization algorithms and introduce a number of trainable parameters in each iteration, we keep the function derived from the finite-iteration method as a whole and adopt DNNs to pre-process the input to this function. Given the universal approximation property, the added neural networks are able to improve the performance of the end-to-end design.
The overall block diagram of the proposed ADU end-to-end design is shown in Fig. \ref{architecture}. In particular, given the perfect downlink CSI at the user side, one encoding DNN is reused by different users to perform CSI quantization. At the BS side, the function derived from the WMMSE beamformer, i.e., $f_{\mathrm{WMMSE}}(\cdot)$, is preserved and $N$ DNNs are deployed to pre-process the input of this function. In Fig. \ref{architecture}, $\mathbf{H}_{i_k}$ denotes the local CSI and $\widetilde{\mathbf{H}}_{i_k}$ denotes the pre-processed local CSI for the $i_k$-th user, respectively. Then, the ADU-based downlink beamforming optimization with limited feedback is reformulated as
\begin{equation} \label{ADU}
\begin{aligned}
\max_{\mathcal{F}(\cdot), \mathcal{H}(\cdot)} ~ & \sum_{i=1}^{M}\sum_{k=1}^{I_i} \log \operatorname{det}(\mathbf{I}+\mathbf{H}_{i_{k}, i} \mathbf{V}_{i_{k}} \mathbf{V}_{i_{k}}^{H} \mathbf{H}_{i_{k}, i}^{H} \\
&(\sum_{m=1, m \neq k}^{I_{i}} \mathbf{H}_{i_{k}, i} \mathbf{V}_{i_{m}} \mathbf{V}_{i_{m}}^{H} \mathbf{H}_{i_{k}, i}^{H}+ \\
&\sum_{j=1, j \neq i}^{M} \sum_{\ell = 1}^{I_j} \mathbf{H}_{i_{k}, j} \mathbf{V}_{j_{\ell}} \mathbf{V}_{j_{\ell}}^{H} \mathbf{H}_{i_{k}, j}^{H}+\sigma_{k}^{2} \mathbf{I})^{-1})\\ 
\text{s.t.} ~ & {\mathbf{q}}_{i_{k}, i}=\mathcal{F}\left(\mathbf{H}_{i_{k}, i}\right), \forall i_k \in \mathcal{I},\forall i \in \mathcal{M}, \\
& \mathbf{V}_{i_{k}}=f_{\mathrm{WMMSE}}\left(\mathcal{H}({\mathbf{q}}_{i})\right), \forall i_k \in \mathcal{I},\forall i \in \mathcal{M},
\end{aligned}
\end{equation}
where $\mathcal{H}(\cdot)$ denotes the pre-processing scheme for WMMSE beamformer. In the following, we shall demonstrate the CSI feedback and downlink beamforming components in detail, respectively. Furthermore, the additional training technique, i.e., the VIB framework, will be introduced.

\subsection{Uplink Feedback Phase}
Prior to data transmission, each user transmits $B$ information bits of quantized CSI back to the associated BS for downlink beamforming. 
Since the channel distribution for different users are independent and identically distributed, one duplicate DNN can be used to encode the CSI at all $N$ users regardless of the number of users in the system. 
An $R$-layer fully-connected DNN is adopted at the users to perform the channel quantization where the feedback bits of user $i_k$ can be written as
\begin{align}\label{eq_NN1}  \nonumber
& \mathbf{q}_{i_k} =\\
&\operatorname{sgn}\left(\mathbf{W}^{(i_k)}_{R}\sigma_{{R-1}}\left( \cdots  \sigma_{1}\left( \mathbf{W}_{1}^{(i_k)} \mathbf{x}_{i_k} + \mathbf{b}_{1}^{(i_k)} \right)  \cdots \right)+\mathbf{b}_{R}^{(i_k)} \right).
\end{align}
Here $\{\mathbf{W}_{r}^{(i_k)},\mathbf{b}_{r}^{(i_k)} \}_{{r} = {1}}^{R}$ denotes the trainable parameters for user $i_k$ and
$\sigma_{r}$ is the activation function for the $r$-th layer. The activation function for the last layer is implemented by the sign function $\operatorname{sgn}(\cdot)$, which is used to guarantee that the output of user side DNN is zero-one bit stream $\mathbf{q}_{i_k}$. 
In \eqref{eq_NN1}, the real representation of $\mathbf{x}_{i_k}$, i.e.,
\begin{equation}
  \mathbf{x}_{i_k} \triangleq \left[\operatorname{flat}(\Re\left( \mathbf{H}_{i_k, j}\right)) , \operatorname{flat}(\Im\left( \mathbf{H}_{i_k, j}\right)) \right]^T,  j \in \mathcal{M},
\end{equation}
is considered as the input of the DNN since most of the existing deep learning libraries only support real-value operations, where $\operatorname{flat}(\cdot)$ denotes the flatten operation that reshapes a matrix into a row vector.

Due to the fact that the gradient of the binarization
neuron is almost zero everywhere, the conventional back-propagation training method
cannot be directly used to train the layers prior to the binarization layer \cite{sohrabi21deep}.
To overcome this issue, a common solution in the machine learning literature is to
approximate the activation function of a binarization layer by another
smooth and differentiable function during training.
One of the most popular approaches is sigmoid-adjusted straight-through (ST), which replaces
the derivative factor with the gradient of the function $2
\operatorname{sigm}(u) -1$, where $\operatorname{sigm}(u) = 1/({1+\exp(-u)})$
is the sigmoid function.  
In this paper, we adopt the sigmoid-adjusted ST with annealing \cite{sohrabi21deep} in the training stage to compute the gradients of the binary layer considered at the last layer of the user side DNN.

\subsection{Downlink Beamforming Phase}
To incorporate the domain knowledge inherent in the iterative optimization algorithm, in the proposed ADU method, we adopt the function derived from the WMMSE beamformer, i.e., $f_{\mathrm{WMMSE}}(\cdot)$, as the basis for augmentation and implement DNNs to pre-process the input of this function.
It is shown in \cite{ma2021learn} that for a suboptimal handcrafted algorithm, there exists a calibrated input which achieves a higher system performance. Specifically, there exists a pre-processed input $\widetilde{\mathbf{H}}_{i_k, j}$ that achieves higher performance when it is fed into the limited-iteration WMMSE algorithm. That is, the sum-rate achieved by $f_\mathrm{WMMSE}(\widetilde{\mathbf{H}}_{i_k, j})$ is higher than that achieved by $f_\mathrm{WMMSE}(\mathbf{H}_{i_k, j})$. However, even if the calibrated input exists, it is typically unknown and hard to analytically characterize. In this case, the powerful learning capabilities of DNNs are leveraged to approximate the complicated mapping $\widetilde{\mathbf{H}}_{i_k, j} = \mathcal{P}(\mathbf{H}_{i_k, j})$, where $\mathcal{P}(\cdot)$ denotes the pre-processing procedure.


Let $\mathcal{Q}(\cdot)$ denote the downlink CSI reconstruction mapping that maps the feedback bits $\mathbf{q}_{i_k}$ into the reconstructed channel $\mathbf{H}_{i_k, j}$. Note that, instead of considering $\mathcal{Q}(\cdot)$ and $P(\cdot)$ as two separate modules and using two different neural networks to approximate them, we directly learn the composite mapping $\mathcal{H}(\cdot) \triangleq \mathcal{Q}(\mathcal{P}(\cdot))$. By doing so, the proposed ADU method integrates the pre-processing operation for the unfolded WMMSE algorithm with the channel reconstruction operation, which provides the opportunity to achieve a global optimum for the joint design task.
However, the input and output dimensions of the parameterized mapping $\mathcal{H}(\cdot)$ are proportional to both the number of transmit antennas and the number of users, which are large numbers in MU-MIMO systems. Employing giant and unstructured neural networks does not incorporate the uniqueness of Problem \eqref{ADU} and is not applicable when user density is large.



In this paper, we develop a customized learning model by identifying the permutation equivariance property of Problem \eqref{ADU}. We model the considered multi-cell MU-MIMO network as a directed graph with edge and node features. In particular, each user is modeled as one node in the graph and forms a complete graph. The node feature tensor $\mathbf{Z} \in \mathbb{C}^{N \times N_r \times N_t}$ is given by $\mathbf{Z}_{(i_k, :, :)} = \mathbf{H}_{i_k, i}$. The adjacency feature tensor $\mathbf{A} \in \mathbb{C}^{N \times N \times N_r \times N_t}$ is given by $\mathbf{A}_{(i_k, j_\ell, :, :)} = \mathbf{H}_{i_k, j}$. Then \eqref{ADU} can be rewritten as an optimization problem over a graph, given by
\begin{equation} \label{ADU_PE}
\begin{aligned}
\max_{\mathcal{F}(\cdot), \mathcal{H}(\cdot)} ~ & \sum_{i=1}^{M}\sum_{k=1}^{I_i} \log \operatorname{det}(\mathbf{I}+\mathbf{Z}_{(i_k, :, :)} \mathbf{V}_{i_{k}} \mathbf{V}_{i_{k}}^{H} \mathbf{Z}_{(i_k, :, :)}^{H} \\
&(\sum_{m=1, m \neq k}^{I_{i}} \mathbf{Z}_{(i_k, :, :)} \mathbf{V}_{i_{m}} \mathbf{V}_{i_{m}}^{H} \mathbf{Z}_{(i_k, :, :)}^{H}+ \\
&\sum_{j=1, j \neq i}^{M} \sum_{\ell = 1}^{I_j} \mathbf{A}_{(i_k, j_\ell, :, :)} \mathbf{V}_{j_{\ell}} \mathbf{V}_{j_{\ell}}^{H} \mathbf{A}_{(i_k, j_\ell, :, :)}^{H}+\sigma_{k}^{2} \mathbf{I})^{-1})\\ 
\text{s.t.} ~ & {\mathbf{q}}_{i_{k}, i}=\mathcal{F}\left(\mathbf{Z}_{(i_k, :, :)}\right), \forall i_k \in \mathcal{I},\forall i \in \mathcal{M}, \\
& \mathbf{V}_{i_{k}}=f_{\mathrm{WMMSE}}\left(\mathcal{H}({\mathbf{q}}_{i})\right), \forall i_k \in \mathcal{I},\forall i \in \mathcal{M},
\end{aligned}
\end{equation}
It is shown in \cite[Proposition 3]{Shen21Graph} that the permutation equivariance property is universal for optimization over a graph. This indicates that for the downlink beamforming task in multi-cell MU-MIMO systems, it is the elements in $\mathbf{H}_{i_k, j}$ rather than the ordering of different channel matrices that count when maximizing the sum-rate. This allows us to share trainable weights among different users. Therefore, we develop $N$ duplicate DNNs that share the common trainable parameters for $N$ users in the cellular systems to approximate the mapping $\mathcal{H}(\cdot)$. The input and output dimensions of each DNN are then reduced by a factor of $N$, which is independent of the number of users. This makes the proposed ADU-based method scalable for large number of users and does not increase the computational and storage cost during training.


\subsection{Variational Information Bottleneck}
In limited feedback multi-cell MU-MIMO systems, there is a natural trade-off between the system performance and the feedback overhead. 
Specifically, if more feedback bits are transmitted, the BS will get a more accurate channel information, which leads to higher spectral efficiency. 
Therefore, transmitting sufficient but minimal channel information that is essential for the multi-user beamforming task is the key design point especially when the number of feedback bits is strictly limited. 
The IB framework proposed in \cite{tishby99information} has been applied to investigate the data fitting and generalization trade-off in deep learning. The IB framework maximizes the mutual information between the latent representation and the output label to obtain high accuracy, while minimizing the mutual information between the representation and the input data sample to improve generalization. Such a trade-off between preserving the significant information and identifying a compact representation is consistent with the limited feedback system design and thus will be adopted as an additional design principle in our study.

For the considered Problem \eqref{ADU}, and different from methods proposed in \cite{sohrabi21deep, Guo21Cellular, Hu21Unfold} that use the negative sum-rate as the training loss function, 
the training loss function of the proposed ADU is given by
\begin{equation}\label{loss_IB}
L = -\sum_{i=1}^{M}\sum_{k=1}^{I_i} R_{i_k} + \gamma I (\mathbf{q}, \mathbf{H}),
\end{equation}
where $I (\cdot, \cdot)$ represents the mutual information function and $\gamma > 0$ denotes the regularization parameter controlling the trade-off. The first term in \eqref{loss_IB} promotes the DNN to boost system performance, i.e., maximize the system sum-rate, while the second term in \eqref{loss_IB} forces the DNN to find the information encompassed in the CSI as compressive as possible.

However, the computation of mutual information term for high-dimensional data with unknown distributions is challenging. This is because the empirical estimate for the probability distribution requires that the number of sampling to increase exponentially with the dimension \cite{Shao22Edge}. In this paper, we adopt a VIB technique \cite{alemi2017variational} to deal with the mutual information computation of the loss function in \eqref{loss_IB}. The VIB framework introduces a set of approximating densities to the intractable distribution. By definition, the mutual information $I (\mathbf{q}, \mathbf{H})$ is expressed as
\begin{equation}
\begin{aligned}
I (\mathbf{q}, \mathbf{H}) & = \int d \mathbf{q} d \mathbf{H} p(\mathbf{q}, \mathbf{H}) \log \frac{p(\mathbf{q} \mid \mathbf{H})}{p(\mathbf{q})} \\
& =\int d \mathbf{q} d \mathbf{H} p(\mathbf{H}, \mathbf{q}) \log p(\mathbf{q} \mid \mathbf{H})-\int d \mathbf{q} p(\mathbf{q}) \log p(\mathbf{q})
\end{aligned}
\end{equation}

Computing the marginal distribution of $\mathbf{q}$, i.e., $p(\mathbf{q}) = \int d\mathbf{H} p(\mathbf{q}|\mathbf{H})p(\mathbf{H})$ is difficult. Let $r(\mathbf{q})$ be a variational approximation to this marginal. 
Since $\operatorname{KL}[p(\mathbf{q}), r(\mathbf{q})] \geq 0$ and $\int d\mathbf{q} p(\mathbf{q}) \log p(\mathbf{q}) \geq \int d\mathbf{q} p(\mathbf{q}) \log r(\mathbf{q})$, we have the following upper bound
\begin{equation}\label{VIB}
I(\mathbf{q}, \mathbf{H}) \leq \int d \mathbf{H} d \mathbf{q} p(\mathbf{H}) p(\mathbf{q} \mid \mathbf{H}) \log \frac{p(\mathbf{q} \mid \mathbf{H})}{r(\mathbf{q})}.
\end{equation}
By further applying the re-parameterization trick and Monte Carlo sampling \cite{alemi2017variational, Shao22Edge}, we are able to obtain an unbiased estimate of the gradient and hence optimize the objective using stochastic gradient descent.

\section{Simulation Results} \label{simulation}
In this section, we demonstrate the performance of the proposed ADU-based end-to-end design in MU-MIMO limited feedback systems. 

\subsection{Simulation Setup}
A cellular network with 9 cells is simulated. At the center of each cell, a BS is deployed to synchronously serve $K$ users which are located uniformly and randomly within the cell range $r \in [R_\mathrm{min},R_\mathrm{max}]$, where $R_\mathrm{min} = 0.01$ km and $R_\mathrm{max} = 1$ km are the inner space and half cell-to-cell distance, respectively. The small-scale fading is simulated to be Rayleigh distributed. According to the LTE standard, the large-scale fading is modeled as $\beta= 120.9+37.6 \log_{10}(d)+10 \log_{10}(z)$ dB, where $z$ is a log-normal random variable with standard deviation being 8 dB, and $d$ is the transmitter-to-receiver distance (km). The AWGN power $\sigma^2$ is -114 dBm and the emitting power constraints are 35 dBm.

The user-wise shared DNN adopted in the proposed method has 4 fully-connected layers with 1024, 512, 256, and $B$ neurons in each layer, respectively. The BS-side DNN has 4 fully-connected layers with 512, 2048, 2048, and $2 N_\mathrm{t} N_\mathrm{r}$ neurons in each layer, respectively. 4-iteration WMMSE is unfolded at the BS side. The objective function in (9) is used as the unsupervised loss. We train the neural network for 200 epochs using the Adam optimizer with a minibatch size of 1024 and a learning rate of 0.001. There are in total 204,800 training samples and 1000 test samples. After each dense layer, the batch normalization layer is leveraged to stabilize convergence and the rectified linear unit (ReLU) is utilized as the activation function in the hidden layers.

\subsection{Performance Comparison}
To illustrate the effectiveness of the proposed neural calibration end-to-end design, we adopt three benchmarks for comparisons:
\begin{itemize}
\item \textbf{Fully Data-Driven}: The black-box DNNs in \cite{sohrabi21deep} are adopted to map the downlink CSI to the feedback bits at users and map the feedback bits to the downlink beamforming matrix at the BSs.
\item \textbf{Codebook-based}: Conventional random vector quantization is used to construct the downlink channel quantization codebook.
\item \textbf{GCN-WMMSE w/ DNNCR}: Conventional block-by-block scheme is used. For CSI feedback, a fully-connected auto-encoder and auto-decoder are used for channel quantization and reconstruction, respectively. For downlink beamforming, deep unfolding method in \cite{schynol2022coordinated} is implemented.
\end{itemize}

\begin{figure}[t] 
\centering
\includegraphics[height=5.0cm]{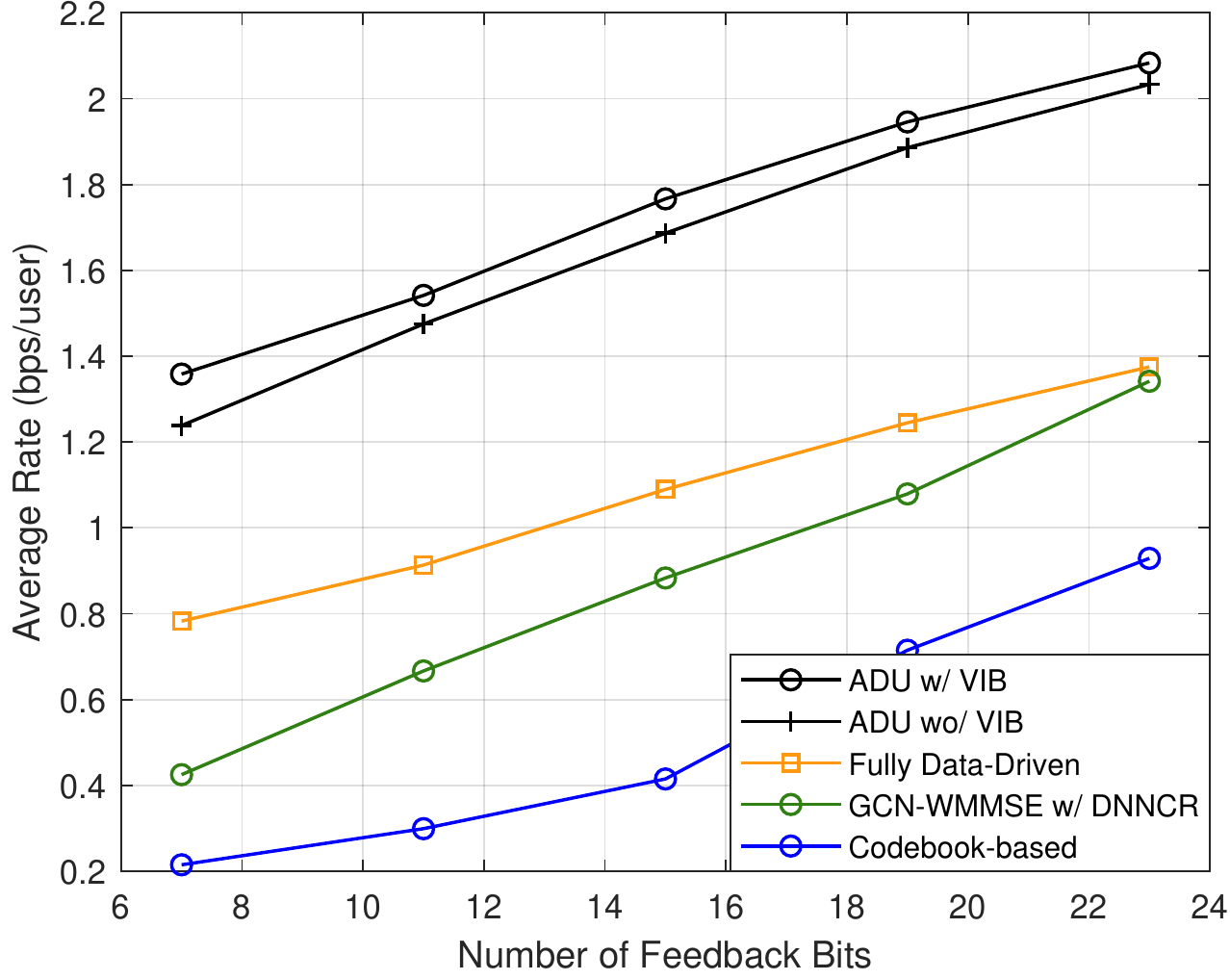} 
\caption{Averaged rate achieved by different methods when $N_\mathrm{t} = 64$, $N_\mathrm{r} = 2$, and $K=4$.} 
\label{figureK4} 
\end{figure}

Fig. \ref{figureK4} plots the average data rate achieved by the proposed scheme and the three baseline methods versus the number of feedback bits. It is demonstrated that the proposed ADU-based design outperforms all the other baselines. 
The average rate achieved by the proposed ADU method is significantly higher than that of the GCN-WMMSE w/ DNNCR over the whole regime, which shows the effectiveness of ADU-based joint design over the conventional block-by-block unfolding scheme. Furthermore, it indicates that simply cascading two modules cannot achieve a satisfactory performance even if both are deep learning-assisted.
It is also demonstrated that when the number of feedback bits increases, the performance of the fully data-driven method gets restricted and the performance gain over the block-by-block scheme vanishes when $B = 23$. This implies that without domain knowlege, the conventional black-box deep learning method suffers from poor scalability. 
Besides, the ADU with the VIB technique further improves the average data rate, indicating that the effectiveness of forcing DNNs to find compressive and informative representation.

\begin{figure}[t] 
\centering
\includegraphics[height=5.0cm]{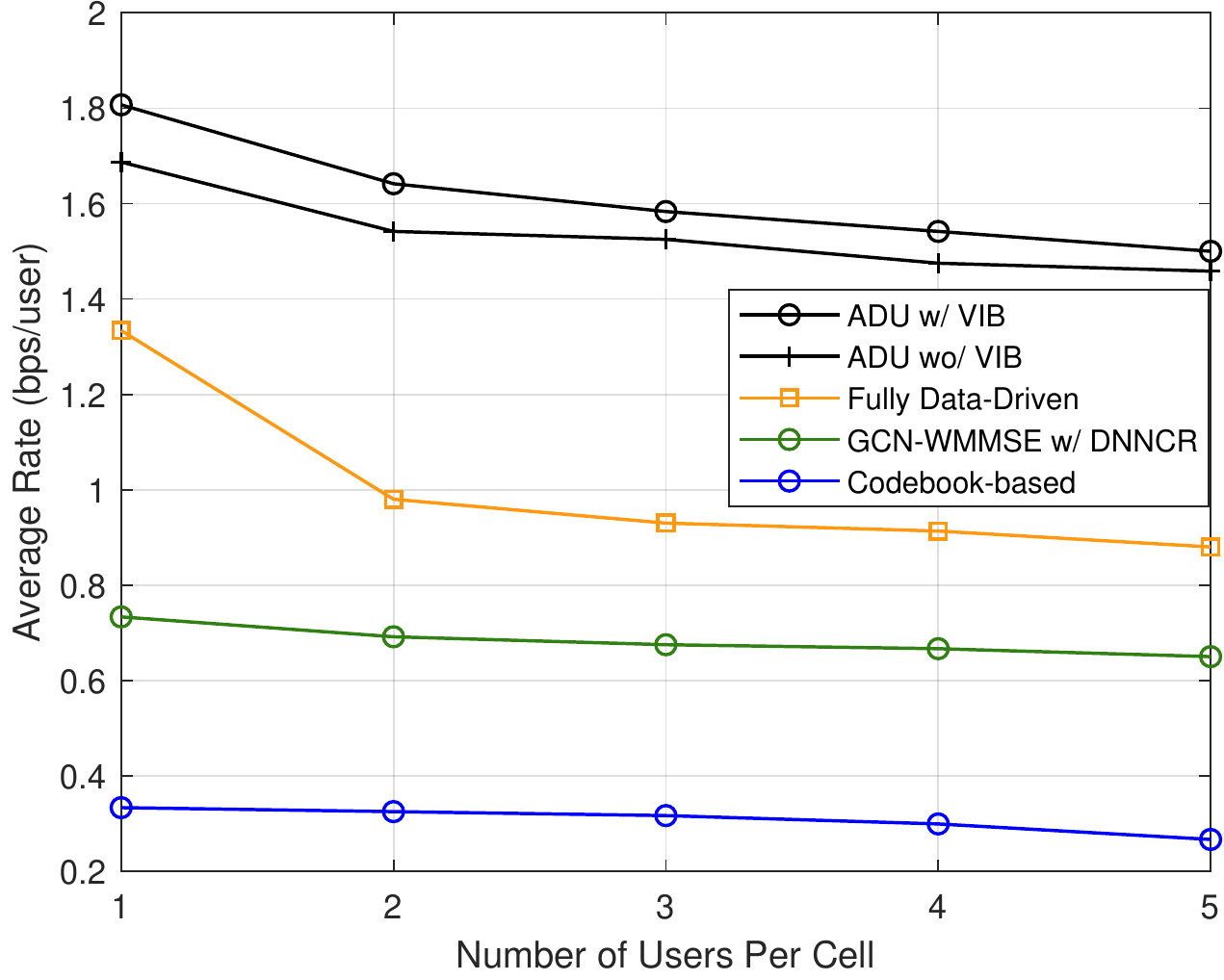} 
\caption{Averaged rate achieved by different methods when $N_\mathrm{t} = 64$, $N_\mathrm{r} = 2$, and $B = 11$.} 
\label{figureK} 
\end{figure}

In Fig. \ref{figureK}, we demonstrate the system average data rate versus the number of users per cell. As can be observed in Fig. \ref{figureK}, the fully data-driven method entails a prominent performance loss when the number of users per cell increases, showing its limited capability of managing interference. The proposed ADU-based design significantly outperforms the conventional block-by-block method and fully data-driven method for all investigated values of $K$. 
This verifies the superiority of the proposed design in terms of both the data rate and scalability in wireless networks where users are densely distributed. Note that as $K$ increases, the approximation of the probability distribution in VIB becomes less accurate, resulting in less performance gain compared with the one without VIB when $K$ is large.

\begin{figure}[t] 
\centering
\includegraphics[height=5.0cm]{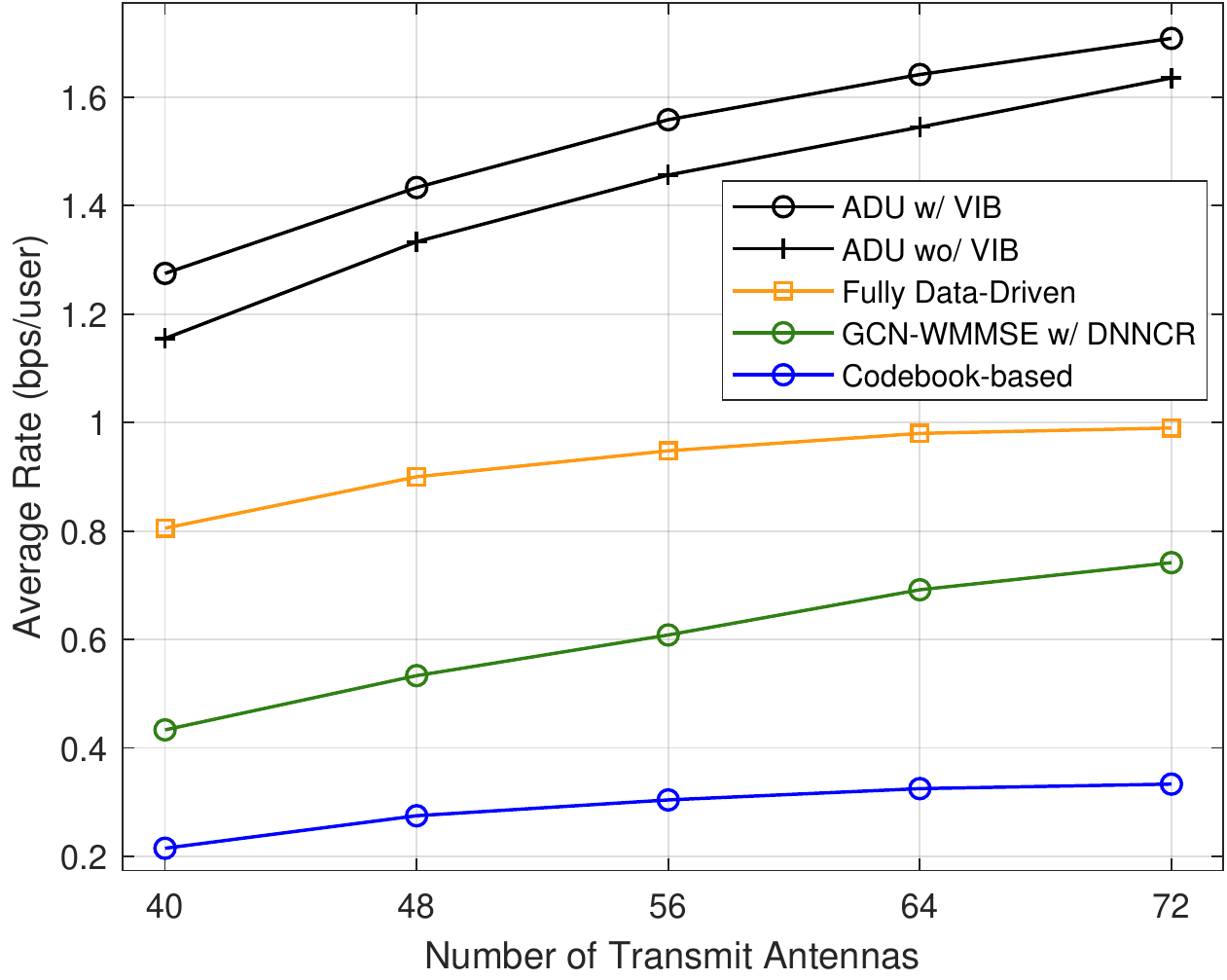} 
\caption{Averaged rate achieved by different methods when $K = 2$, $N_\mathrm{r} = 2$, and $B = 11$.} 
\label{figureM} 
\end{figure}

Fig. \ref{figureM} plots the average data rate versus the number of transmit antennas at the BSs. Both the proposed ADU method and the GCN-WMMSE w/ DNNCR baseline combine domain knowledge with deep learning, and thus achieves a similar performance trend when $N_\mathrm{t}$ increases. However, the fully data-driven method suffers from performance loss for large-scale transmit antennas, due to its black-box nature. It is demonstrated that the proposed ADU-based design outperforms all the other baselines, especially when $N_\mathrm{t}$ is large. This confirms the superiority of the proposed ADU design in terms of average data rate for multi-cell MU massive MIMO systems.

%

\section{Conclusions}
In this paper, we developed an ADU-based method for downlink beamforming in limited feedback multi-cell MU-MIMO systems. In contrast to existing deep unfolding methods that unroll an iterative algorithm and introduce a number of trainable parameters in each layer, in the proposed method, the function derived from the conventional fixed-iteration numerical method is kept intact and DNNs are leveraged to pre-process its input for a better performance.
Simulation results clearly demonstrated that the proposed ADU-based end-to-end design achieves an excellent performance in large-scale multi-cell MU-MIMO systems with a limited number of feedback bits.

\bibliographystyle{IEEEtran}
\bibliography{IEEEabrv,CSIQBF_V2}

\begin{thebibliography}{10}
\providecommand{\url}[1]{#1}
\csname url@samestyle\endcsname
\providecommand{\newblock}{\relax}
\providecommand{\bibinfo}[2]{#2}
\providecommand{\BIBentrySTDinterwordspacing}{\spaceskip=0pt\relax}
\providecommand{\BIBentryALTinterwordstretchfactor}{4}
\providecommand{\BIBentryALTinterwordspacing}{\spaceskip=\fontdimen2\font plus
\BIBentryALTinterwordstretchfactor\fontdimen3\font minus
  \fontdimen4\font\relax}
\providecommand{\BIBforeignlanguage}[2]{{%
\expandafter\ifx\csname l@#1\endcsname\relax
\typeout{** WARNING: IEEEtran.bst: No hyphenation pattern has been}%
\typeout{** loaded for the language `#1'. Using the pattern for}%
\typeout{** the default language instead.}%
\else
\language=\csname l@#1\endcsname
\fi
#2}}
\providecommand{\BIBdecl}{\relax}
\BIBdecl

\bibitem{boccardi2014five}
F.~Boccardi, R.~W. Heath, A.~Lozano, T.~L. Marzetta, and P.~Popovski, ``Five
  disruptive technology directions for {5G},'' \emph{IEEE Commun. Mag.},
  vol.~52, no.~2, pp. 74--80, 2014.

\bibitem{KBL226G}
K.~B. Letaief, Y.~Shi, J.~Lu, and J.~Lu, ``Edge artificial intelligence for
  {6G}: Vision, enabling technologies, and applications,'' \emph{IEEE J. Sel.
  Areas Commun.}, vol.~40, no.~1, pp. 5--36, Jan. 2022.

\bibitem{Jun09TWC}
J.~Zhang, R.~Chen, J.~G. Andrews, A.~Ghosh, and R.~W. Heath, ``Networked {MIMO}
  with clustered linear precoding,'' \emph{IEEE Trans. Wireless Commun.},
  vol.~8, no.~4, pp. 1910--1921, Apr. 2009.

\bibitem{Vincent14CS}
X.~Rao and V.~K.~N. Lau, ``Distributed compressive {CSIT} estimation and
  feedback for {FDD} multi-user massive {MIMO} systems,'' \emph{IEEE Trans.
  Signal Process.}, vol.~62, no.~12, pp. 3261--3271, June 2014.

\bibitem{sohrabi21deep}
F.~Sohrabi, K.~M. Attiah, and W.~Yu, ``Deep learning for distributed channel
  feedback and multiuser precoding in {FDD} massive {MIMO},'' \emph{IEEE Trans.
  Wireless Commun.}, vol.~20, no.~7, pp. 4044--4057, July 2021.

\bibitem{Guo21Cellular}
J.~Guo, C.-K. Wen, and S.~Jin, ``Deep learning-based {CSI} feedback for
  beamforming in single- and multi-cell massive {MIMO} systems,'' \emph{IEEE J.
  Sel. Areas Commun.}, vol.~39, no.~7, pp. 1872--1884, July 2021.

\bibitem{HDCSI}
J.-C. Shen, J.~Zhang, K.-C. Chen, and K.~B. Letaief, ``High-dimensional {CSI}
  acquisition in massive {MIMO}: Sparsity-inspired approaches,'' \emph{IEEE
  Syst. J.}, vol.~11, no.~1, pp. 32--40, 2017.

\bibitem{Shi11WMMSE}
Q.~{Shi}, M.~{Razaviyayn}, Z.~{Luo}, and C.~{He}, ``An iteratively weighted
  {MMSE} approach to distributed sum-utility maximization for a {MIMO}
  interfering broadcast channel,'' \emph{IEEE Trans. Signal Process.}, vol.~59,
  no.~9, pp. 4331--4340, Sept. 2011.

\bibitem{Yifan21NC}
Y.~Ma, Y.~Shen, X.~Yu, J.~Zhang, S.~Song, and K.~B. Letaief, ``Neural
  calibration for scalable beamforming in {FDD} massive {MIMO} with implicit
  channel estimation,'' in \emph{2021 IEEE Global Commun. Conf. (GLOBECOM)},
  Madrid, Spain, Dec. 2021, pp. 1--6.

\bibitem{He20Model}
H.~He, C.-K. Wen, S.~Jin, and G.~Y. Li, ``Model-driven deep learning for {MIMO}
  detection,'' \emph{IEEE Trans. Signal Process.}, vol.~68, pp. 1702--1715,
  Feb. 2020.

\bibitem{Hu21Unfold}
Q.~{Hu}, Y.~{Cai}, Q.~{Shi}, K.~{Xu}, G.~{Yu}, and Z.~{Ding}, ``Iterative
  algorithm induced deep-unfolding neural networks: Precoding design for
  multiuser {MIMO} systems,'' \emph{IEEE Trans. Wireless Commun.}, vol.~20,
  no.~2, pp. 1394--1410, Feb. 2021.

\bibitem{ma2021learn}
Y.~Ma, Y.~Shen, X.~Yu, J.~Zhang, S.~Song, and K.~B. Letaief, ``Learn to
  communicate with neural calibration: Scalability and generalization,''
  \emph{arXiv preprint arXiv:2110.00272}, 2021.

\bibitem{Shen21Graph}
Y.~{Shen}, Y.~{Shi}, J.~{Zhang}, and K.~B. {Letaief}, ``Graph neural networks
  for scalable radio resource management: Architecture design and theoretical
  analysis,'' \emph{IEEE J. Sel. Areas Commun.}, vol.~39, no.~1, pp. 101--115,
  Jan. 2021.

\bibitem{tishby99information}
\BIBentryALTinterwordspacing
N.~Tishby, F.~C. Pereira, and W.~Bialek, ``The information bottleneck method,''
  in \emph{Proc. Annu. Allerton Conf. Commun. Control Comput.}, Monticello, IL,
  USA, Oct. 2000, pp. 368--377. [Online]. Available:
  \url{https://arxiv.org/abs/physics/0004057}
\BIBentrySTDinterwordspacing

\bibitem{Shao22Edge}
J.~Shao, Y.~Mao, and J.~Zhang, ``Learning task-oriented communication for edge
  inference: An information bottleneck approach,'' \emph{IEEE J. Sel. Areas
  Commun.}, vol.~40, no.~1, pp. 197--211, Jan. 2022.

\bibitem{alemi2017variational}
A.~A. Alemi, I.~Fischer, J.~V. Dillon, and K.~Murphy, ``Deep variational
  information bottleneck,'' in \emph{Proc. Int. Conf. Learn. Represent.},
  Toulon, France, Apr. 2017.

\bibitem{schynol2022coordinated}
L.~Schynol and M.~Pesavento, ``Coordinated sum-rate maximization in multicell
  {MU-MIMO} with deep unrolling,'' \emph{arXiv preprint arXiv:2202.10371}, Feb.
  2022.

\end{thebibliography}

\end{document}